\documentstyle[12pt]{article}
\def\QATOP#1#2{{#1 \atop #2}}
\date{}
\begin{document}

\title{The Quantum Skyrmion in Representations of General Dimension}
\author{A. Acus$^1$ E. Norvai\v{s}as$^1$ and D.O. Riska$^2$}
\maketitle

\centerline{\it $^1$Institute of Theoretical Physics and Astronomy,
Vilnius, 2600 Lithuania}

\centerline{\it $^2$Department of Physics, University of Helsinki,
00014 Finland}

\vspace{1cm}

\begin{abstract}
The representations of general dimension are constructed for the
$SU(2)$ Skyrme model, treated quantum mechanically {\it ab initio. }
This quantum Skyrme model has a negative mass term
correction, that is not present in the classical Hamiltonian. 
The magnitude of the quantum mechanical mass correction
increases with the dimension of the representation of the $SU(2)$ 
group. In the case of a 5-dimensional representation it is possible to obtain
satisfactory predictions for the nucleon
mass with the empirical value for the pion decay constant.
\end{abstract}

\newpage
{\bf 1. Introduction}
\vspace{0.5cm}

The $SU(2)$ version of Skyrme's topological soliton model for the
baryons [1, 2] is conventionally described with field operators that
belong to the fundamental 2-dimensional representation of the $SU(2)$
group. At the classical level the predictions for the baryon
observables and phenomenology turn out to be independent of the 
dimension of the
representation used for the group in the case of the original verion
of the Skyrme model [3]. This situation changes when 
the
Lagrangian density of the Skyrme model is treated quantum 
mechanically
{\it ab initio.} In this case, as will be shown below, the negative 
purely
quantum mechanical mass correction that arises in the systematic
quantization [4] of the model, is representation dependent, and grows in
magnitude and significance with the dimension of the representation. \\

In this paper the quantum mechanical treatment of the Skyrme model in
a representation of arbitrary dimension will be developed. The
theoretical formalism builds on that developed for the classical
treatment of the Skyrme model in a general representation in ref. [3].
A quantum mechanical mass
formula for the baryon states is derived. In addition the
expressions for the Noether and anomalous current operators are
derived. Finally we study the dependence of the predictions for the
phenomenological baryon structure parameters on the dimension of the
presentation numerically, and show that the quantum
mechanical treatment makes it possible to obtain satisfactory
predictions for the baryon masses with the empirical value
of the pion decay constant if a 5-dimensional representation is employed.\\

This paper is divided into 5 sections. In section 2 the
classical treatment of the Skyrme model in a representation of 
general
dimension [3] is reviewed. In section 3 the quantum mechanical treatment
of the Skyrme
model is developed. In section 4 the Noether currents of the
Lagrangian density are derived, along with the expressions for the
magnetic moments of the nucleons and the $\Delta_{33}$ resonances
as well as the axial coupling 
constant of the nucleon.
Numerical results for these observables are given in section 5. 
Section 6 contains a concluding discussion.\\

\vspace{1cm}

{\bf 2. The classical skyrmion in a general representation.}
\vspace{0.5cm}

The Skyrme model is based on a Lagrangian density for a unitary
field $U(\vec r, t)$ that belongs to an
irreducible representation of 
the  
 $SU(2)$ group. In a general 
irreducible representation it is convenient to express the unitary field $U$
in terms of three unconstrained
Euler angles ${\bf \alpha}=(\alpha ^1,\alpha ^2,\alpha ^3)$ as
$$
U({\bf x},t)=D^j({\bf \alpha }({\bf x},t)).\eqno(2.1) 
$$
The elements of the matrices $D^j$ are the
Wigner D-functions, where $(2j+1)$ is the dimension of the
$SU(2)$ representation. The Euler angles ${\bf \alpha }$ then form 
the 
dynamical
variables of the theory.\\

The Skyrme model is defined by the chirally 
symmetric
Lagrangian density 
$$
{\cal L[}U({\bf x},t){\cal ]}=-{\frac{f_\pi ^2}4}{\rm Tr}\{R_\mu 
R^\mu \}+{%
\frac 1{32e^2}}{\rm Tr}\{[R_\mu ,R_\nu ]^2\},\eqno(2.2) 
$$
where the ''right'' current $R_\mu $ is defined as 
$$
R_\mu =(\partial _\mu U)U^{\dagger },\eqno(2.3)$$
and $f_\pi $ (the pion decay constant) and $e$ are parameters. As was 
shown
in [3] the classical Lagrangian density depends on the dimension of
the representation $j$ only
through the overall scalar factor%
$$
N={2\over 3}j(j+1)(2j+1).\eqno(2.4) 
$$
which can be incorporated in the parameters by a renormalization.
As a consequence the equations of motion for the dynamical variables 
${\bf %
\alpha }$ are independent of the dimension of the representation.\\

The trace of a bilinear combination of two generators of the group 
$\hat
J_a,\hat J_b$ depends on
the dimension of the representation as 
$$
{\rm Tr}\langle jm|\hat J_a\hat J_b|jm^{\prime }\rangle =(-)^a{\frac 16}%
j(j+1)(2j+1)\delta _{a,-b}.\eqno(2.5) 
$$
The commutator relations for the generators are
$$
[\hat J_a,\,\hat J_b]=\left[ 
\begin{array}{ccc}
1 & 1 & 1 \\ 
a & b & c 
\end{array}
\right] \hat J_c.\eqno(2.6) 
$$
Here the factor on the r.h.s. is the Clebsch-Gordan coefficient 
$(1a1b|1c)$%
, in a more transparent notation. The components of the operators
$\hat J_a$ are defined above as 
$\hat J_{\pm }=-J_{\pm 1}/\sqrt{2}$ and $\hat J_0=-J_0/\sqrt{2}$.

The ''spherically symmetric'' hedgehog ansatz in a general 
representation is
invariant under the combined spatial and isospin rotation
$$
i\left[ {\bf x\times \nabla }\right] _aU({\bf x)}+\sqrt{2}\left[ 
J_a,U({\bf %
x)}\right] =0,\eqno(2.7) 
$$
where circular components are used for both the vector and isovector. 
The
solution of (2.7) is the generalization of the usual hedgehog ansatz 
$$
e^{i\vec \tau \cdot {\bf x}F(r)}\Longrightarrow U_0({\bf x)}=\exp 
{\bigl[ -}i%
\sqrt{2}\hat {{J}}{_a\cdot \hat x^aF(r)\bigr] }.\eqno(2.8) 
$$
Here the circular coordinates of the unit vector $\hat {{\bf x}}$ 
are
defined as
$$
\begin{array}{l}
\hat x^{+1}=-\frac 1{
\sqrt{2}}\left( \hat x_1-i\hat x_2\right) =-\frac 1{\sqrt{2}}\sin 
\theta
e^{-i\varphi }=-\hat x_{-1,} \\ \hat x^0=\hat x_3=\cos \theta =\hat 
x_0, \\ 
\hat x^{-1}=\frac 1{\sqrt{2}}\left( \hat x_1+i\hat x_2\right) =\frac 
1{\sqrt{%
2}}\sin \theta e^{i\varphi }=-\hat x_{+1.} 
\end{array}
\eqno(2.9) 
$$
The generalized hedgehog ansatz $U_0({\bf x})=D^j\bigl( {\bf \beta 
}\left( 
{\bf x}\right) \bigr) $ can be expressed in terms of Euler angles as
$$
\begin{array}{l}
\beta ^1( 
{\bf x})=\varphi -\arctan (\cos \vartheta \,\tan F(r))-\pi /2, \\ 
\beta ^2( 
{\bf x})=-2\arcsin (\sin \vartheta \,\sin F(r)), \\ \beta ^3({\bf x}%
)=-\varphi -\arctan (\cos \vartheta \,\tan F(r))+\pi /2. 
\end{array}
\eqno(2.10) 
$$
Here the angles $\varphi$ and $\vartheta$ are the polar angles that
the define the direction of the unit vector $\hat x$.\\

With the hedgehog ansatz (2.8) the Lagrangian density (2.2) reduces 
to the
following simple form%
$$
{\cal L}(F(r))=-{\frac 43}j(j+1)(2j+1)\bigg\{{\frac{f_\pi ^2}4}%
\Bigl(F^{^{\prime }2}+{\frac 2{r^2}}\sin ^2\!F\Bigr) 
$$

$$
+{\frac 1{16e^2}}{\frac{\sin ^2\!F}{r^2}}\Bigl(2F^{^{\prime 
}2}+{\frac{\sin
^2\!F}{r^2}}\Bigr)\bigg\}.\eqno(2.11) 
$$

The requirement that the soliton mass be stationary yields the same
differential equation for the chiral angle $F(r)$ as in [2]. The 
overall
factor $N$ (2.4) does not affect the solution and hence the classical 
soliton is
independent of the dimension of the representation. \\

After the renormalization the hedgehog mass in any representation $j$ 
has the 
form%
$$
M(F)=\frac{f_\pi }e\tilde M(F)= 
$$
$$
2\pi\frac{f_\pi }e \int d\widetilde{r}\widetilde{r}^2\left[ F^{\prime 
2}+%
\frac{\sin ^2F}{\widetilde{r}^2}\left( 2+2F^{\prime 2}+\frac{\sin 
^2F}{%
\widetilde{r}^2}\right) \right] ,\eqno(2.12) 
$$
where the dimensionless variable 
$\widetilde{r}$ is defined as $\widetilde{r}=ef_\pi r$ [2]. 
Variation of the mass leads to the standard differential equation 
$$
F^{\prime \prime }+2F^{\prime \prime }\frac{\sin 
^2F}{\widetilde{r}^2}%
+F^{\prime 2}\frac{\sin 2F}{\widetilde{r}^2}+\frac 2{\widetilde{r}%
^2}F^{\prime }-\frac{\sin 2F}{\widetilde{r}^2}-\frac{\sin 2F\sin 
^2F}{%
\widetilde{r}^4}=0\eqno(2.13) 
$$
for the chiral angle $F(r)$.\\

For the hedgehog solution the baryon density takes the form%
$$
B^0=\frac 1{24\pi ^2N}\epsilon ^{0\nu \beta \gamma }{\rm Tr}\,R_\nu
\,R_\beta \,R_\gamma =-\frac 1{2\pi ^2}\frac{\sin 
^2F}{r^2}F^{^{\prime }},%
\eqno(2.14) 
$$
The renormalization factor $N$ ensures that the lowest nonvanishing 
baryon 
number is $%
B=1 $ for the hedgehog in all representations.  \\

\vspace{1cm}

{\bf 3. Quantization of skyrmion in collective coordinate approach.}
\vspace{0.5cm}

The quantization of Skyrme model in a general dimension is a bit
intricate [3]. Following Adkins {\it et.al. }[2] we shall employ
collective rotational coordinates to separate the variables which 
depend 
on the time
and spatial coordinates:
$$
U({\bf x},{\bf q}(t))=A\left( {\bf q}(t)\right) U_0({\bf 
x})A^{\dagger
}\left( {\bf q}(t)\right). \eqno(3.1) 
$$
The set of three real, independent parameters ${\bf q}%
(t)=(q^1(t),q^2(t),q^3(t))$ are quantum variables (skyrmion rotation 
Euler
angles). In a general representation the unconstrained variables 
${\bf q}(t)$ are more convenient 
than the four constrained
Euler-Rodrigues parameters used in [2]. We shall 
consider
the Skyrme Lagrangian (2.2) quantum mechanically {\it ab initio.} The
generalized coordinates ${\bf q}(t)$ and velocities ${\bf \dot q(}t)$
then satisfy the commutation relations [5]:
$$
[\dot q^a,\,q^b]=-if^{ab}({\bf q}).\eqno(3.2) 
$$
Here the tensor $f^{ab}({\bf q)}$ is a function of generalized 
coordinates $%
{\bf q}$ only, the explicit form of which is determined after the
quantization condition has been imposed. The tensor $f^{ab}$ is 
symmetric
with respect to interchange of the indices $a$ and $b$ as a 
consequence of
the commutation relation $[q^a,\,q^b]=0$. The commutator relation 
between a
generalized velocity component $\dot q^a$ and arbitrary function 
$G({\bf q})$
is given by 
$$
[\dot q_a,\,G({\bf q})]=-i\sum_rf^{ar}({\bf q}){\frac \partial 
{\partial q^r}%
}G({\bf q}).\eqno(3.3)$$

After making the substitution (3.1) into the Lagrangian density (2.2) 
the dependence of the Lagrangian on the generalized
velocities can be expressed as
$$
L({\bf \dot q,q,}F)=\frac 1N\int {\cal L(}{\bf x,q}(t),F(r){\cal 
)}r^2\sin
\vartheta drd\vartheta d\varphi = 
$$
$$
-\frac 14a(F)\dot q^\alpha g({\bf q})_{\alpha \beta }\dot q^\beta 
+[(\dot q)^0 {\rm -order\ term }]\eqno(3.4) 
$$
Here $a(F)$ is defined as the constant.
$$
a(F)=\frac 1{e^3f_\pi }\widetilde{a}(F)=\frac 1{e^3f_\pi }\frac{8\pi 
}3\int d%
\widetilde{r}\widetilde{r}^2\sin ^2F\left[ 1+F^{\prime 2}+\frac{\sin 
^2F}{%
\widetilde{r}^2}\right] ,\eqno(3.5) 
$$
The 3$\times $3 metric tensor $g({\bf q})_{\alpha \beta }$ is defined 
as the
scalar product of a set of functions  
$C_\alpha ^{(m)}({\bf q})$ [3] as
$$
g({\bf q})_{\alpha \beta }=\sum_m(-)^mC_\alpha ^{(m)}C_\beta
^{(-m)}=\sum_m(-)^mC_\alpha ^{^{\prime }(m)}C_\beta ^{^{\prime 
}(-m)}= 
$$
$$
-2\delta _{\alpha \beta }-2(\delta _{\alpha 1}\delta _{\beta 
3}+\delta
_{\alpha 3}\delta _{\beta 1})\cos q^2,\eqno(3.6) 
$$
where the functions $C_\alpha^{'(m)}$ are defined as
$$
C_\alpha ^{(m)}({\bf q})=\sum_mD_{m,m^{\prime }}^1({\bf q})C_\alpha 
^{\prime
(m)}({\bf q}).\eqno(3.7) 
$$
The orthogonality relations for the functions
$C_\alpha^{(m)}$ are 
$$
\sum_mC_\alpha ^{(m)}C_{(m)}^\beta =\sum_mC_\alpha ^{\prime
(m)}C_{(m)}^{\prime \beta }=\delta _{\alpha ,\beta },\eqno(3.8) 
$$
$$
\sum_\alpha C_\alpha ^{(m)}C_{(n)}^\alpha =\sum_\alpha C_\alpha 
^{\prime
(m)}C_{(n)}^{\prime \alpha }=\delta _{m,n}.\eqno(3.9) 
$$

The appropriate definition for the canonical momentum $p_\alpha $, 
which is conjugate 
to $%
q^\alpha $, is%
$$
p_\alpha ({\bf \dot q,q,}F)=\frac{\partial L({\bf \dot 
q,q,}F)}{\partial
\dot q^\alpha }=-\frac 14a(F)\{\dot q^\beta ,g({\bf q})_{\beta \alpha 
}\},%
\eqno(3.10) 
$$
where the curly bracket denotes an anticommutator. The canonical 
commutation
relations%
$$
\left[ p_\alpha ({\bf \dot q,q,}F),q^\beta \right] =-i\delta _{\alpha 
\beta
},\eqno(3.11) 
$$
then yield the following explicit form for the functions (3.2) 
$f^{ab}({\bf q})$ 
$$
f^{ab}({\bf q})=-\frac 2{a(F)}g_{\alpha \beta }^{-1}({\bf 
q}).\eqno(3.12) 
$$

Because of the nonlinearity of the Skyrme model the canonical momenta 
defined in this way do not necessarily 
satisfy
the relation $[p_\alpha ,p_\beta ]=0$. As shown
in [5], there does however exist a local transformation of the set of
variables ${\bf q}$, 
which
makes it possible to satisfy these relations.  
Define the angular momentum operator 
$$
\hat J_a^{\prime }=-\frac i2\left\{ p_r,C_{\left( -a\right) }^{\prime 
r}(%
{\bf q})\right\} =(-)^a\frac{ia(F)}4\left\{ \dot q^r,C_r^{\prime 
\left(
-a\right) }({\bf q})\right\} ,\eqno(3.13) 
$$
which satisfies the commutation relations (2.5).
The operator $\hat J_a^{\prime }$ is then a ''right
rotation'' generating matrix $D^j({\bf q)}$:
$$
\left[ \hat J_a^{\prime },D_{m,m^{\prime }}^l({\bf q)}\right] 
=-\left\langle
l,m^{\prime }+a\left| \hat J_a\right| l,m^{\prime }\right\rangle
D_{m,m^{\prime }+a}^l({\bf q)},\eqno(3.14) 
$$
and 
$$
\hat J_a=-\frac i2\left\{ p_r,C_{\left( -a\right) }^r({\bf 
q})\right\} =(-)^a%
\frac{ia(F)}4\left\{ \dot q^r,C_r^{\left( -a\right) }({\bf 
q})\right\} ,%
\eqno(3.15) 
$$
is a ''left rotation'' generating matrix $D^j({\bf q)}$:
$$
\left[ \hat J_a,D_{m,m^{\prime }}^l({\bf q)}\right] =\left\langle 
l,m\left|
\hat J_a\right| l,m-a\right\rangle D_{m-a,m^{\prime }}^l({\bf 
q).}\eqno(3.16)
$$
Some lengthy manipulation yields the following explicit 
form
for the Lagrangian:
$$
L({\bf \dot q,q,}F)=-M(F)-\Delta M_j(F)+\frac 1{a(F)}\hat {J^{\prime 
}}^2= 
$$
$$
-M(F)-\Delta M_j(F)+\frac 1{a(F)}\hat J^2,\eqno(3.17) 
$$
where%
$$
\Delta M_j(F)=e^3f_\pi \cdot \Delta \tilde{M_j}(F)=e^3f_\pi 
\frac{-2\pi 
}{5\widetilde{a}^2(F)}\int d\widetilde{r}\widetilde{r}^2\sin 
^2F\times 
$$
$$
\Bigl[5+2(2j-1)(2j+3)\sin ^2F+[2j(j+1)+1]\frac{\sin ^2F}{\tilde r^2} 
$$
$$
+[8j(j+1)-1]F^{\prime 2}-2(2j-1)(2j+3)F^{\prime 2}\sin 
^2F\Bigr].\eqno(3.18) 
$$
The corresponding Hamilton operator is then
$$
H_j(F)=M(F)+\Delta M_j(F)+\frac 1{a(F)}\hat {J^{\prime 
}}^2=M(F)+\Delta
M_j(F)+\frac 1{a(F)}\hat J^2.\eqno(3.19) 
$$

The most important feature of this result is that the quantum 
correction $\Delta M_j(F)$ is negative definite and that it depends
explicitly on the dimension of the 
representation of the 
$SU(2)$ group. This term is lost in the usual semiclassical treatment 
of
the Skyrme model even in the fundamental representation of $SU(2)$,
because that ignores the commutation relations (3.2).
In the numerical work reported below we shall have to treat this
quantum mass correction as a perturbation, 
and use the classical equation of motion for the chiral angle
$F(r)$ (2.13) that is obtained by variation of the classical mass
expression (2.12).
This implies that the quantum skyrmion 
is considered as a rigid rotating classical skyrmion, where
the collective variables describe the spinning mode of the model. 
This perturbative treatment of the quantum correction is motivated
by the fact that the equation
of motion that would be obtained by requiring the quantum
mass expression (3.18) to be stationary has physically
acceptable solutions only in a very narrow parameter window
[6].

For the Hamiltonian (3.19) are the normalized state 
vectors
with fixed spin and isospin $\ell $ 
$$
\left| \QATOP{\ell }{m,m^{\prime }}\right\rangle =\frac{\sqrt{2\ell 
+1}}{%
4\pi }D_{m,m^{\prime }}^\ell ({\bf q})\left| 0\right\rangle, 
\eqno(3.20) 
$$
with the eigenvalues  
$$
H(j,\ell ,F)=M(F)+\Delta M_j(F)+\frac{\ell (\ell 
+1)}{2a(F)}.\eqno(3.21) 
$$

\vspace{1cm}

{\bf 4. The Noether currents.}
\vspace{0.5cm}

The Lagrangian density of the Skyrme model is invariant under left 
and 
right
transformations of the unitary field $U$. The corresponding Noether 
currents can be expressed in
terms of the collective coordinates (3.1). The vector and axial 
Noether 
currents that are
associated with the transformations

$$
U(x)\stackrel{V(A)}{\longrightarrow }\left( 1-i2\sqrt{2}\omega ^a\hat
J_a\right) U(x)\left( 1+(-)i2\sqrt{2}\omega ^a\hat J_a\right) 
\eqno(4.1) 
$$
are nevertheless simpler and directly related to physical 
observables. The 
factor $-2%
\sqrt{2}$ before the generators is introduced so that
the transformation (4.1) for $j=1/2$ matches
the infinitesimal transformation in [2]. 
The 
Noether
currents are operators in terms of the generalized collective 
coordinates 
$q$
and the generalized angular momentum operator $\hat J^{\prime }$ 
(3.13). 
The
explicit expression for the vector current density is%
$$
\hat V_b^a=\frac{\partial {\cal L}_V}{\partial \left( \nabla ^b\omega
_a\right) }=\frac{4\sqrt{2}}3j(j+1)(2j+1)\frac{\sin ^2F}r\Biggl({\rm 
i}%
\biggl\{ f_\pi ^2+\frac 1{e^2}\Bigl(
F^{\prime }{}^2 
$$
$$
+\frac{\sin ^2F}{r^2}-\frac{2(2j-1)(2j+3)+5}{4\cdot 5\cdot 
a^2(F)}\sin ^2F%
\Bigr) \biggr\}
\left[ 
\begin{array}{ccc}
1 & 1 & 1 \\ 
u & s & b 
\end{array}
\right] D_{a,s}^1({\bf q})\hat x_u 
$$
$$
-\frac{\sin ^2F}{\sqrt{2}\cdot e^2\cdot a^2(F)}(-)^s\Bigl\{\left[ 
\hat
J^{\prime }\times \hat {{\bf x}}\right] _{-s}D_{a,s}^1({\bf q})\left[ 
\left[
\hat J^{\prime }\times \hat {{\bf x}}\right] \times \hat {{\bf 
x}}\right] _b 
$$
$$
+\left[ \left[ \hat J^{\prime }\times \hat {{\bf x}}\right] \times 
\hat {%
{\bf x}}\right] _bD_{a,s}^1({\bf q})\left[ \hat J^{\prime }\times 
\hat {{\bf %
x}}\right] _{-s}\Bigr\}\Biggr).\eqno(4.2) 
$$
Here $\nabla ^k$ is a circular component of the gradient  
operator. The
indexes $a$ and $b$ denote isospin and spin components. The time 
(charge) component
of the vector current density becomes%
$$
\hat V_t^a=\frac{\partial {\cal L}_V}{\partial \left( \partial 
_0\omega
_a\right) }=\frac{4\sqrt{2}}{3\cdot a(F)}j(j+1)(2j+1)\sin ^2F\left[ 
f_\pi
+\frac 1{e^2}\left( F^{\prime 2}+\frac{\sin ^2F}{r^2}\right) \right] 
\times 
$$
$$
(-)^s\left\{ D_{a,-s}^1({\bf q})\hat J_s^{\prime }-D_{a,-s}^1({\bf 
q})\hat
x_s(\hat J^{\prime }\cdot \hat {{\bf x}})\right\}. \eqno(4.3) 
$$
The explicit expression for the axial current density takes the form%
$$
\hat A_b^a=\frac{\partial {\cal L}_A}{\partial \left( \nabla ^b\omega
_a\right) }=\frac 23j(j+1)(2j+1)\Biggl( \Biggl\{ f_\pi ^2\frac{\sin 
2F}%
r+\frac 1{e^2}\frac{\sin 2F}r\biggl( F^{\prime }{}^2+\frac{\sin 
^2F}{r^2} 
$$
$$
-\frac{\sin ^2F}{4\cdot a^2(F)}\biggr) \Biggr\}
D_{a,b}^1({\bf q})+\Biggl\{ f_\pi ^2\Bigl(2F^{\prime }-\frac{\sin 
2F}%
r\Bigr)-\frac 1{e^2}\biggl(F^{\prime }{}^2\frac{\sin 2F}r-4F^{\prime 
}\frac{%
\sin ^2F}{r^2} 
$$
$$
+\frac{\sin ^2F\sin 2F}{r^3}-\frac{\sin ^2F\sin 2F}{4\cdot 
a^2(F)\cdot r}%
\Bigr)\Biggr\} (-)^sD_{a,s}^1({\bf q})\hat x_{-s}\hat 
x_b-\frac{2F^{\prime
}\sin ^2F}{e^2\cdot a^2(F)}\times 
$$
$$
(-)^s\left\{ D_{a,s}^1({\bf q})\hat x_{-s}\hat J^{\prime }{}^2+\hat
J^{\prime }{}^2D_{a,s}^1({\bf q})\hat x_{-s}-2D_{a,s}^1({\bf q})\hat
x_{-s}(\hat J^{\prime }\cdot \hat {{\bf x}})(\hat J^{\prime }\cdot 
\hat {%
{\bf x}})\right\} \hat x_b\times 
$$
$$
-\frac{\sin ^2F\sin 2F}{e^2\cdot a^2(F)\cdot r}(-)^s\left\{ \left[ 
\left[
\hat J^{\prime }\times \hat {{\bf x}}\right] \times \hat {{\bf 
x}}\right]
_{-s}D_{a,s}^1({\bf q})\left[ \left[ \hat J^{\prime }\times \hat 
{{\bf x}%
}\right] \times \hat {{\bf x}}\right] _b\right. 
$$
$$
\left. +\left[ \left[ \hat J^{\prime }\times \hat {{\bf x}}\right] 
\times
\hat {{\bf x}}\right] _bD_{a,s}^1({\bf q})\left[ \left[ \hat 
J^{\prime
}\times \hat {{\bf x}}\right] \times \hat {{\bf x}}\right] 
_{-s}\right\} 
\Biggr)
\eqno(4.4) 
$$
The operators (4.2),(4.3) and (4.4) are well defined for all 
representations $j$
of the classical soliton and for fixed spin and isospin $l$ of the 
quantum 
skyrmion.
The new terms which are absent in the semiclassical case are those
that have the factor 
$a^2(F)$ in the 
denominator.

The conserved topological current density in Skyrme model is the 
baryon
current density, the components of which are%
$$
{\cal B}_a{\cal (}{\bf x,}F(r){\cal )}=\frac 1{\sqrt{2}\pi ^2a(F)\ 
r}\sin
^2F\cdot F^{\prime }\left[ \hat{J}^{\prime }\times \hat{{\bf 
x}}\right] _a.%
\eqno(4.5) 
$$

The matrix elements of the third component of the corresponding 
isoscalar magnetic 
moment operator 
have the form%
$$
\left\langle \QATOP{\ell }{m_tm_s}\right| \left[ \mu _{I=0}\right]
_3\left| \QATOP{\ell }{m_tm_s}\right\rangle =\left\langle \QATOP{\ell 
}{%
m_{t\,}m_s}\right| \frac 12\int {\rm d}^3xr\left[ \hat {{\bf 
x}}\times {\cal %
B}\right] _0\left| \QATOP{\ell }{m_{t\,}m_s}\right\rangle = 
$$
$$
\frac{\left[ l(l+1)\right] ^{1/2}e}{3\cdot \tilde a(F)f_\pi 
}\left\langle
\tilde r_{I=0}^2\right\rangle \left[ 
\begin{array}{ccc}
\ell & 1 & \ell \\ 
m_s & 0 & m_s 
\end{array}
\right] ,\eqno(4.6) 
$$
where the mean square radius is given as
$$
\left\langle \tilde r_{I=0}^2\right\rangle =-\frac 2\pi \int \tilde 
r^2\sin
^2F\cdot F^{\prime }d\tilde r,\eqno(4.7) 
$$
and $\tilde{a}$ is defined in eq. (3.5).\\

The matrix elements of the third component of the isovector part of 
magnetic
moment operator that is obtained from the isovector current (4.3) 
have 
the form 
$$
\left\langle \QATOP{\ell }{m_tm_s}\right| \left[ \mu _{I=1}\right]
_3\left| \QATOP{\ell }{m_tm_s}\right\rangle =\left\langle \QATOP{\ell 
}{%
m_{t\,}m_s}\right| \frac 12\int {\rm d}^3x\cdot r\left[ \hat {{\bf 
x}}\times
\hat V^3\right] _0\left| \QATOP{\ell }{m_{t\,}m_s}\right\rangle = 
$$
$$
\left[ \frac{\tilde a(F)}{e^3\cdot f_\pi }+\frac{8\pi \cdot e}{3\cdot 
\tilde
a^2(F)\cdot f_\pi }\int d\tilde r\cdot \tilde r^2\sin ^4F\right. 
\left( 1-%
\frac{(2j-1)(2j+3)}{2\cdot 5}\right. 
$$
$$
-\frac{l(l+1)}3\left. \left. +\frac{(-)^{2l}}2{\Biggl[\frac{%
5l(l+1)(2l-1)(2l+1)(2l+3)}{2\cdot 3}\Biggr]}^{1/2}\left\{ 
\begin{array}{ccc}
1 & 2 & 1 \\ 
l & l & l 
\end{array}
\right\} \right) \right] \times 
$$
$$
\left[ 
\begin{array}{ccc}
\ell & 1 & \ell \\ 
m_s & 0 & m_s 
\end{array}
\right] \left[ 
\begin{array}{ccc}
\ell & 1 & \ell \\ 
m_t & 0 & m_t 
\end{array}
\right] ,\eqno(4.8). 
$$
where the symbol in the curly brackets is a $6j$ coefficient.

From the axial 
current density (4.4) we obtain the axial coupling constant $g_A$ of
the nucleon as 
$$
g_A=-3\left\langle \QATOP{1/2 }{1/2,1/2}\right| \int {\rm 
d}^3xA_0^0\left| 
\QATOP{1/2 }{1/2,1/2}\right\rangle =\frac 
1{e^2}\tilde{g_1}(F)-\frac{\pi
^2e^2}{3\cdot \tilde{a}^2(F)}\left\langle 
\tilde{r}_{I=0}^2\right\rangle,
\eqno(4.9) 
$$
where%
$$
\tilde{g_1}(F)=\frac{4\pi }3\int d\tilde{r}\Bigl( 
\tilde{r}^2F^{\prime }+%
\tilde{r}\sin 2F+\tilde{r}\sin 2F\cdot F^{\prime 2} 
$$
$$
+2\sin ^2F\cdot F^{\prime }+\frac{\sin ^2F}{\tilde{r}}\sin 2F\Bigr) 
.%
\eqno(4.10) 
$$

\vspace{1cm}
\newpage
{\bf 5. The static properties of the nucleon and the $\Delta_{33}$ 
resonance}
\vspace{0.5cm}

The $I=J=\ell =1/2$ and $I=J=\ell =3/2$ skyrmions are to be 
identified with 
the nucleons and the $\Delta_{33}$ resonances. As in ref. 
[2] we determine the two parameters in the Lagrangian density
(2.2) so that their masses 
take their empirical values. The expressions 
for
the nucleon and $\Delta_{33}$ masses are
$$
m_N=\frac{f_\pi }e\tilde{M}(F)+e^3f_\pi \cdot \Delta 
\tilde{M}_j(F)+\frac{%
e^3f_\pi }{2\cdot \tilde{a}(F)}\frac 34,\eqno(5.1) 
$$
$$
m_\Delta =\frac{f_\pi }e\tilde{M}(F)+e^3f_\pi \cdot \Delta 
\tilde{M}_j(F)+%
\frac{e^3f_\pi }{2\cdot \tilde{a}(F)}\frac{15}4.\eqno(5.2) 
$$
In the evaluation of these two masses numerically we employ the 
chiral
angle $F(r)$, which is obtained by solving the classical equation of
motion that is given by the requirement that the classical mass 
(2.12)
by stationary. The corresponding values for the Lagrangian parameters
are given in Table 1 for different values of the dimension $(2j+1)$ 
of
the $SU(2)$ representation.\\

In the table we also include the predicted values for the other 
static
nucleon properties, as well as the original predictions obtained in
ref. [2] for the classical Skyrme model. In the case of the
fundamental representation $j=1/2$ the numerical importance of the
quantum correction is small, as was to be expected. For larger values
of $j$ the quantum corrections become increasingly important. The key
qualitative feature is that the quantum mass correction $\Delta
M_j(F)$ is negative, and hence it becomes possible to reproduce the
empirical nucleon and $\Delta_{33}$ mass values with increasingly
realistic values of the pion decay constant $f_\pi$. This reaches its
empirical value 93 MeV for a representation of dimension 5. There is
an accompanying improvement of the numerical value for the axial
coupling constant $g_A$. \\

In the case of the isoscalar radius $r_0$ of the baryon, there is
however no reduction of the difference between the predicted 
and the empirical value
with increasing dimension of the representation. The same is true for the
magnetic moments. The predicted value for the ratio of the proton and
neutron magnetic moments deteriorates slowly with increasing
dimension of the representation.\\

\vspace{1cm}

{\bf 6. Discussion}
\vspace{0.5cm}

Once the Skyrme model is treated consistently quantum mechanically 
{\it ab
initio} the dimension of the representation of the $SU(2)$ group
becomes a significant additional model parameter. When the dimension
of the representation in increased to 5 from the value 2 for the
fundamental representation, it becomes possible to obtain 
satisfactory
values for the masses of the nucleon and the $\Delta_{33}$ resonance
with a value for the pion decay constant, which is very close to the
empirical value (89.4 MeV vz. 93 MeV). There is unfortunately no
comparable gain in quality of the predictions for the baryon magnetic
moments, which deteriorate slowly with increasing dimension of the
representation. The value of the axial coupling constant does on the
other hand improve, but stays below 1 for representations of 
reasonably
low dimension. The fact that the axial coupling constant remains low
is a natural consequence of the vanishing axial charge commutator in
the Skyrme model [7,8].\\

Note that the perturbative treatment used here for the quantum
skyrmion breaks down when the dimension of the representation grows so
large that the negative quantum mass correction becomes of the same
order of magnitude as or larger than the classical skyrmion mass. This
feature is clearly related to the fact that the equation of motion for
the quantum skyrmion has physically acceptable solutions only in a narrow
parameter window [6].\\

The numerical value of the quantum mass correction $\Delta M_j(F)$
(3.18) is of the order 100 MeV in the fundamental representation,
but it rapidly increases in magnitude as the dimension of the
representation grows. For a representation of dimension 5 it is large
enough to cancel the $\sim 500$ MeV overprediction of the nucleon 
mass
that obtains when the empirical value for the pion decay constant is
employed in the classical Skyrme model. It is interesting to note 
that
it plays a similar role to the (negative) Casimir correction to the
Skyrmion energy considered in ref. [9].

\vspace{1cm}
{\bf Acknowledgment} \vspace{0.5cm}

The research of A.A and E.N 
was made possible in part by the Long-Term Research Grants 
NN {\bf LA5000} and {\bf LHU100 } from the International Science Foundation.
\vspace{1cm}

{\bf References} \vspace{0.5cm}

\begin{enumerate}
\item  T.H.R. Skyrme, Proc. Roy. Soc. {\bf A260} (1961) 127

\item  G.S. Adkins, C.R. Nappi and E. Witten, Nucl. Phys. {\bf B228} 
(1983)
552

\item  E. Norvai\v sas and D.O. Riska, Physica Scripta. {\bf 50} 
(1994) 634

\item  E.M. Nyman and D.O. Riska, Rept. Prog. Phys. {\bf 53} (1990) 
1137

\item  K. Fujii, A. Kobushkin, K. Sato and N. Toyota, Phys. Rev. {\bf 

D35}
(1987) 1896

\item A. Kostyuk, A. Kobushkin, N. Chepilko and T. Okazaki,
Yad. Fiz. {\bf 58 } (1995) 1488

\item N. Kaiser, U. Vogl and W. Weise, Nucl. Phys. A {\bf 484} (1988)
493, {\bf 490} (1988) 602

\item M. Kirchbach and D.O. Riska, Nuovo. Cim. {\bf A104} (1991) 1837

\item B. Moussallam, Phys. Lett. {\bf B272} (1991) 196

\item Particle Data Group, Phys. Rev. {\bf D50} (1994) 1173
\item  A. Bosshard et al., Phys. Rev. {\bf D44}
(1991) 1962

\end{enumerate}
\vspace{2cm}
\newpage 

Table 1. The predicted static baryon observables as obtained with 
the
quantum Skyrme model for representations of different dimension. The
first column (ANW) are the predictions for the classical Skyrme model
given in ref. [2]. The empirical results [10,11] are listed in the last 
column.
\vspace{1cm}

\begin{tabular}{|l|l|l|l|l|l|l|l|} \hline
  &  ANW &  $j=1/2$ &
$j=1$ & $j=3/2$ & $j=2$ & $j=5/2$ &
Exp.\\ \hline
  $m_N$       &  input &  input &  input      &  input
&  input      &  input &$\hphantom{0}939$ MeV. \\
  $m_\Delta $ &  input &  input &  input      &  input
&  input      &  input &$1232$ MeV. \\
  $f_\pi $    &$\hphantom{0}64.5$  &$\hphantom{0}72.1$ &%
$\hphantom{0}76.4$ &$\hphantom{0}82.2$
&$\hphantom{0}89.4$ &$\hphantom{0}98.0$  &%
$\hphantom{00}93$ MeV. \\
 $e $    &$\hphantom{00}5.45$    &$\hphantom{00}5.23$&%
$\hphantom{00}5.15$ &$\hphantom{00}5.03$
&$\hphantom{00}4.89$ &$\hphantom{00}4.74$ & \\
 $r_0 $    &  $\hphantom{00}0.59$   &$\hphantom{00}0.55$ &%
$\hphantom{00}0.53$ &$\hphantom{00}0.51$
&$\hphantom{00}0.48$      & $\hphantom{00}0.45$  &%
$\hphantom{000}0.72$ fm.   \\
 $\mu_p $    & $\hphantom{00}1.87 $   &$\hphantom{00}1.90$ &%
$\hphantom{00}1.84$ &$\hphantom{00}1.78$
&$\hphantom{00}1.71$ & $\hphantom{00}1.64$ &$\hphantom{000}2.79$   \\
 $\mu_n $    & $ -1.31 $ &$-1.42$ &$-1.40$&$-1.37$
&$-1.35 $     &$-1.33$ &$\hphantom{0}-1.91$    \\
 $g_A $    &$\hphantom{00}0.61 $   &$\hphantom{00}0.65$ &%
$\hphantom{00}0.68$ &$\hphantom{00}0.71$
&$\hphantom{00}0.76$  & $\hphantom{00}0.80$  &$\hphantom{000}1.23$    
\\
$\mu_{\Delta ^{++}}$    &   &$\hphantom{00}3.70$ &%
 $\hphantom{m}3.58$ &$ \hphantom{m}3.44$
&$ \hphantom{m}3.29$  & $\hphantom{m}3.15$  &$\hphantom{000}4.52$ 
 \\
$\mu_{\Delta ^{+}}$    &  &$\hphantom{m}1.71$ &
 $\hphantom{m}1.64$ &$ \hphantom{m}1.55$
&$ \hphantom{m}1.46$  & $\hphantom{m}1.37$  &$\hphantom{000}? $   
\\
$\mu_{\Delta ^{0}}$    &  &$-0.28$ &
 $-0.31$ &$ -0.34$
&$-0.38$  &$-0.42$  &$\hphantom{000}?$    \\
$\mu_{\Delta ^{-}}$    &  &$-2.27$ &
 $-2.25$ &$ -2.23$
&$-2.21$  &$-2.20$  &$\hphantom{000}?$    \\
\hline
\end{tabular}

\vspace{2cm}

\end{document}